\documentclass[twocolumn,showpacs,aps,amsmath,amssymb]{revtex4}
\usepackage{graphicx}

\newcommand{\Sec}[1]{Sec.\;\ref{#1}}

\newcommand{\be}{\begin{equation}}
\newcommand{\ee}{\end{equation}}
\newcommand{\bea}{\begin{eqnarray}}
\newcommand{\eea}{\end{eqnarray}}
\newcommand{\bsube}{\begin{subequations}}
\newcommand{\esube}{\end{subequations}}
\newcommand{\Eq}[1]{Eq.\,(\ref{#1})}

\newcommand{\Fig}[1]{Fig.\,\ref{#1}}

\newcommand{\ind}{{\sf n}}

\newcommand{\rhonup}  {\rho_{\sf n}^{{ }_{\{+\}}}}
\newcommand{\rhondown}{\rho_{\sf n}^{{ }_{\{-\}}}}
\newcommand{\spup}  {\left\vert \uparrow   \right\rangle}
\newcommand{\spdown}{\left\vert \downarrow \right\rangle}
\newcommand{\spupdn}{\left\vert d \right\rangle}
\newcommand{\spzero}{\left\vert 0 \right\rangle}
\newcommand{\dup}{_\uparrow}
\newcommand{\ddn}{_\downarrow}

\begin{document}

\title{Dynamic Coulomb blockade in single--lead quantum dots}

\author{Xiao Zheng} \email{chxzheng@ust.hk}
\author{Jinshuang Jin}
\author{YiJing Yan} \email{yyan@ust.hk}

\affiliation{Department of Chemistry, Hong Kong University
   of Science and Technology, Kowloon, Hong Kong}
%

\date{submitted on 27~July~2008}

\begin{abstract}

We investigate transient dynamic response of an Anderson impurity
quantum dot to a family of ramp--up driving voltage applied to the
single coupling lead. Transient current is calculated based on a
hierarchical equations of motion formalism for open dissipative
systems [J.\ Chem.\ Phys.\ {\bf 128}, 234703 (2008)].
In the nonlinear response and nonadiabatic charging regime, characteristic
resonance features of transient response current reveal distinctly
and faithfully the energetic configuration of the quantum dot.
We also discuss and comment on both the physical and numerical
aspects of the theoretical formalism used in this work.

\end{abstract}



\maketitle

\section{Introduction} \label{intro}

Dynamics of open electronic systems far from equilibrium has
attracted considerable interest in the past few years, since the
dynamic processes are closely related to and thus can be used to
explore the system properties of interest. It is well known that
electron--electron interaction plays a significant role under
realistic experimental conditions \cite{Lu03422}. Therefore,
understanding the electronic dynamics in presence of
electron--electron interaction is of fundamental importance to the
emerging field of nanosciences.

\begin{figure}
\begin{center}
\includegraphics[width=0.95\columnwidth]{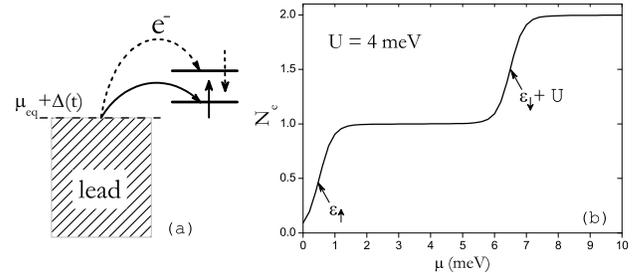}
\caption{ \label{fig1} (a) An interacting QD coupled to a single
lead with a time--dependent chemical potential ${\mu}(t) = \mu_{\rm
eq} + \Delta(t)$. (b) The steady--state number of electrons on QD,
$N_e$, as a function of ${\mu}$. Other parameters adopted are (in
unit of meV): $\epsilon_\uparrow = 0.5$, $\epsilon_\downarrow =
2.5$, $U=4\,$, $T = 0.2$, the QD--lead coupling strength
$\Gamma=0.05$ and bandwidth $W=15$. See \Sec{adiabatic} for details.}
\end{center}
\end{figure}

So far theoretical work has focused mostly on steady or
quasi--steady state dynamics, which however may contain
only partial information on the system of study.
Consider, for example, a quantum dot (QD) as
depicted in \Fig{fig1}(a), with
two Zeeman levels of $\epsilon\dup < \epsilon\ddn$,
in contact with a single electrode.
Under the quasi--steady state
charging condition, the applied voltage raises
the electrode chemical potential ($\mu$) adiabatically.
The first charging electron occupies exclusively
the energetically favorable spin--up
level, while the second one that occupies the spin--closed
double occupation state requires
an excess energy to overcome the static on--dot Coulomb
blockade; see \Fig{fig1}(b).
Apparently, the steady--state study
offers rather incomplete information,
as the single occupation of spin--down state is
dark here.

In this work, we resort to transient
dynamics of the system, by applying a
time--dependent external voltage to the system, \emph{i.e.}, $\mu$
varies explicitly in real time, and analyzing the resulting
nonadiabatic charging process. It is
expected that all transitions within the range
of applied voltage will
be manifested through the system dynamic properties.
Specifically, we consider a single--lead Anderson impurity
QD driven by a family of ramp--up external voltages,
by exploiting the hierarchical equations of motion (HEOM)
theory for open electronic dynamics and
transient current \cite{Jin08234703}.
In \Sec{method} we briefly describe
the theoretical and computational
aspects.
We demonstrate in \Sec{ramp} that the
transient dynamics reveals sensitively and
faithfully the QD energetic configuration.
Section \ref{summary} concludes this work.

\section{Methodology} \label{method}

We implement a HEOM formalism, which is formally exact for dynamics
of an arbitrary non--Markovian dissipative system interacting with
surrounding baths \cite{Jin08234703, Xu05041103, Xu07031107,
Jin07134113}.
From the perspective of quantum dissipation theory, the electronic
dynamics of an open system is mainly characterized by the reduced
system density matrix $\rho(t)\equiv \mbox{tr}_{_{\rm B}}
\rho_{_{\rm T}}(t)$ and the transient current through the coupling
leads. Here $\rho_{_{\rm T}}(t)$ is the total density matrix of the
entire system, and $\mbox{tr}_{_{\rm B}}$ denotes a trace over all
lead degrees of freedom. The final HEOM is cast into a compact form
of \cite{Jin08234703}
\be  \label{eom0}
   \dot\rho_\ind = -[i\mathcal{L} +
    \gamma_\ind(t)] \rho_\ind  + \rhondown + \rhonup.
\ee
Here, $\mathcal{L}\,\cdot \equiv [H(t), \cdot\,]$ is system Liouvillian
(setting $\hbar \equiv 1$). The
basic variables of \Eq{eom0} are $\rho(t)$ and associated auxiliary
density operators (ADOs) $\rho_\ind(t)$, where $\ind$ is an index
set covering all accessible derivatives of the Feynman--Vernon
influence functional \cite{Fey63118}.

For a single--level QD coupled to a single lead, the lead
correlation function can be expanded by an exponential series via a
spectral decomposition scheme \cite{Tan906676, Mei993365, Yan05187}.
In this case $\ind$ involves an $\tilde{n}$--fold combination of
($\sigma, s, m$), which characterizes the exponential series
expansion with $\sigma=\pm$, $s$ the spin, and $m$ the index of
exponents, respectively. Therefore, in \Eq{eom0} $\rho_\ind
\vert_{\tilde{n} = 0} = \rho(t)$ with $\gamma_\ind\vert_{\tilde{n} =
0} = \rhondown\vert_{\tilde{n} = 0} = 0$; $\rho_\ind
\vert_{\tilde{n}
> 0}$ is an ADO at the $\tilde{n}^{{\rm th}}$--tier;
$\gamma_\ind(t)$ collects all the related exponents along with
external voltages, to $\rho_\ind(t)$; and $\rhondown$ and $\rhonup$
are the nearest lower-- and higher--tier counterparts of
$\rho_\ind$, respectively. In particular, the $1^{\rm st}$--tier
ADOs, $\rho_\ind(t)\vert_{\tilde{n}=1}=\rho^\sigma_{sm}(t)$,
determine exclusively the transient current of spin--$s$ through the
lead as \cite{Jin08234703}
\be \label{its}
   I_s(t) \equiv - \mbox{tr}_{_{\rm T}}\! \left[ \hat{N}_s
   \dot{\rho}_{_{\rm T}}(t) \right] = -2 \,\mbox{Im} \sum_{m}
   \mbox{tr}\! \left[ a_s\, \rho^+_{sm}(t)\right].
\ee
$\hat{N}_s$ is the lead total--occupation operator of spin $s$;
$a_s$ is the system annihilation operator of spin $s$; and
$\mbox{tr}_{_{\rm T}}$ and $\mbox{tr}$ denote trace operations over
the total system and reduced system degrees of freedom,
respectively.
Since there is only one coupling lead, the
displacement current at every time $t$
is simply $-\!\sum_s I_s(t)$;
thus the Kirchhoff's current law retains
\cite{Pol05161302, Pol07205308}.
The present HEOM--based quantum transport theory admits
classical geometric capacitors or capacitive coupling to a
gate electrode. The associated displacement current can
then be evaluated readily, leading to
the conservation of total current
\cite{Pol05161302,Pol07205308,Fra03471}. Gauge
invariance \cite{Pol05161302,Pol07205308,Ped9812993} is also
guaranteed at all time, as it can be seen from the form of the
$0^{\rm th}$--tier of the HEOM~\cite{Jin08234703}.

It has been proved \cite{Jin08234703} that for a noninteracting QD,
the HEOM (\ref{eom0}) is exact with a finite terminal tier of
$\tilde{n}_{\rm max} = 2$; while for an interacting QD, in principle
an infinite hierarchy is required for the exact transient dynamics.
In practice a truncation scheme is inevitable. In this work a
straightforward truncation scheme is adopted; that is to set all the
higher--tier ADOs zero: $\rho_\ind \vert_{\tilde{n}>N_{\rm
trun}}=0$, with $N_{\rm trun}$ the preset truncation tier. This scheme
leads to a systematic improvement of results as $N_{\rm trun}$
increases, \emph{i.e.}, by inclusion of higher--tier ADOs, as
verified by extensive numerical tests. For a moderate system--lead
coupling strength $\Gamma< 5\,T$, quantitatively accurate (if not
exact) dynamics is obtained with $N_{\rm trun}=2$, where $T$ is the
temperature in the unit of Boltzmann constant ({\it i.e.}, $k_B=1$).
A higher $N_{\rm trun}$ means a drastic increase in computational
cost. Therefore, considering the tradeoff
between numerical accuracy and computational efforts, all the
following calculations are carried out with $N_{\rm trun}=2$ by
confining the ratio $\Gamma/T$ less than $5$; see \Sec{summary}
for further discussions.

The composite system of interest is described by the Anderson
impurity model~\cite{And6141}, with its Hamiltonian $H_{_{\rm T}}$
expressed as
\be \label{htotal}
   H_{_{\rm T}} = H + h_{_{\rm B}} + H_{_{\rm SB}}.
\ee
$H$ represents the interacting QD of our primary interest:
\be \label{hsys}
   H = \epsilon\dup \hat{n}\dup + \epsilon\ddn
   \hat{n}\ddn + U\, \hat{n}\dup \hat{n}\ddn.
\ee
Here, $\epsilon_s$ is the energy of the spin--$s$ ($\uparrow$ or
$\downarrow$) state, $\hat{n}_s \equiv a^\dag_s a_s$ the system
occupation--number operator of spin $s$, and $U$ the on--dot Coulomb
interaction strength.
In~\Eq{htotal}, $h_{_{\rm B}}=\sum_k \sum_s \epsilon_{ks}\,
\hat{n}_{ks}$ describes the noninteracting coupling lead, with
$\epsilon_{ks}$ being the energy of its single--electron state $k$
of spin $s$, and $\hat{n}_{ks} \equiv d_{ks}^\dag d_{ks}$ the lead
occupation--number operator. The QD--lead coupling is given by
$H_{_{\rm SB}} = \sum_k \sum_s t_{ks}\, d^\dag_{ks }a_s + {\rm
H.c.}$, where $t_{ks}$ is the coupling matrix element between the
lead state $k$ and the QD--level, both with spin $s$.
The widely used Drude model is adopted to capture the coupling lead,
\emph{i.e.}, a Lorentzian spectral density function of $J(\epsilon)
= \Gamma W^2 / (\epsilon^2 + W^2)$, with $W$ characterizing the lead
bandwidth.

We consider explicitly the scenario depicted in \Fig{fig1}(a). A
ramp--up voltage $V(t)$ is applied to the coupling lead at $t=0$,
which excites the QD out of equilibrium. The lead energy levels are
shifted due to the voltage, $\Delta(t) = -eV(t)$, with $e$ being the
elementary charge, and so is the lead chemical potential $\mu(t) =
\mu_{\rm eq} + \Delta(t)$. $\mu_{\rm eq}$ is the equilibrium lead
Fermi energy, which is set to zero hereafter, \emph{i.e.}, $\mu_{\rm
eq} = 0$. Under a ramp--up voltage, $\Delta(t)$ varies linearly with
time until $t=\tau$, and afterwards is kept at a constant amplitude
$\Delta$:
\be \label{ramp0}
   \Delta(t) = \left\{
    \begin{array}{lc}
      \Delta\,t/\tau , \qquad & 0 \le t \le \tau \\
      \ \  \Delta, & t > \tau
    \end{array}
    \right. .
\ee
By tuning the duration parameter $\tau$, the family of ramp--up
voltage covers three distinct regimes: (A) The adiabatic limit at
$\tau \rightarrow\infty$; (B) the intermediate range with a finite
$\tau$; and (C) the instantaneous switch--on limit corresponding to
$\tau \rightarrow 0^+$. These three regimes are individually
explored and elaborated in the subsections of \Sec{ramp}.

The work flow of our numerical procedures are briefly stated here,
while details to be published elsewhere. (a) The lead correlation
function is expanded by exponential functions, and hence establishes
\Eq{eom0}. (b) The equilibrium reduced density matrix and ADOs in
absence of external voltages, $\{\rho_\ind^{\rm eq}\}$, are obtained
by setting both sides of \Eq{eom0} to zero. The linear sparse
problem is solved by the biconjugate gradient method~\cite{Pre92}.
(c) The real--time evolution driven by $V(t)$ is solved via a direct
integration of \Eq{eom0} by employing either the Chebyshev
propagator \cite{Bae049803, Wan07134104} or the $4^{\rm th}$--order
Runge--Kutta algorithm \cite{Pre92}. The outcomes of step (b) are
adopted as the initial conditions for the time evolution in step
(c), \emph{i.e.}, $\rho_\ind(t=0) = \rho_\ind^{\rm eq}$. For every
time $t$, the transient current $I_s(t)$ is evaluated via \Eq{its}.
It normally takes about $1$ days for a typical time evolution to
$100\,$ps with a time step of $0.02\,$ps with a single Core 2 Duo
processor.

\section{Transient response current and its relation to
the system energetic configuration} \label{ramp}

\subsection{The adiabatic limit} \label{adiabatic}

As $\tau$ approaches $\infty$, the infinitely slow application of
external voltage results in quasi--static electronic dynamics for
the open QD system. Consequently, at every time $t$ the QD can be
deemed as in the quasi--equilibrium determined by $\mu$. This
adiabatic charging scenario is introduced in \Sec{intro}, which
provides only partial information on the QD energetic configuration.

The QD physical subspace is completely spanned by the following four
Fock states, \emph{i.e.}, $\spzero$ (vacancy), $\spup$ (single
spin--up occupation), $\spdown$ (single spin--down occupation), and
$\spupdn$ (spin--closed double occupation). The QD gains an energy
of $\epsilon_s$ as it transits from $\spzero$ to $\vert s \rangle$,
or $\epsilon_{\bar{s}} + U$ from $\vert s \rangle$ to $\spupdn$,
respectively. Here $\bar{s}$ labels the spin direction opposite to
$s$. Each value of $\mu$ defines a steady state of zero current for
the QD, together with the number of electrons, $N_e$, residing on it.
Shown in \Fig{fig1}(b) is $N_e$ as a
function of $\mu$ for a specific case of
$\mu_{\rm eq} <\epsilon\dup < \epsilon\ddn$.
The two plateaus
corresponding to quantized $N_e$ are separated by the magnitude of
$U$, and the vertical step prior to each plateau represents a
Fock--state transition which adds one more electron to the QD. The
two steps at $\epsilon\dup$ and $\epsilon\ddn+U$ correspond to the
transitions $\spzero \leftrightarrow \spup$ and $\spup
\leftrightarrow \spupdn$, respectively; while the other two
transitions, $\spzero \leftrightarrow \spdown$ and $\spdown
\leftrightarrow \spupdn$ associated with the excitation energies
$\epsilon\ddn$ and $\epsilon\dup+U$, are missing from \Fig{fig1}(b).
This is because $\mu$ is considered to increase adiabatically;
therefore, the first electron occupying the QD
exclusively enters spin--up level that is energetically more
favorable, and consequently, the second electron occupation
requires an excess energy of
$U$ to overcome the static Coulomb blockade.
We note that
steady states offer rather incomplete information on the system of
interest. To acquire further knowledge, it is inevitable to go
beyond the adiabatic charging limit and explore the transient
regime.

\subsection{The finite switch--on duration regime} \label{finite}

\begin{figure}
\begin{center}
\includegraphics[width=0.95\columnwidth]{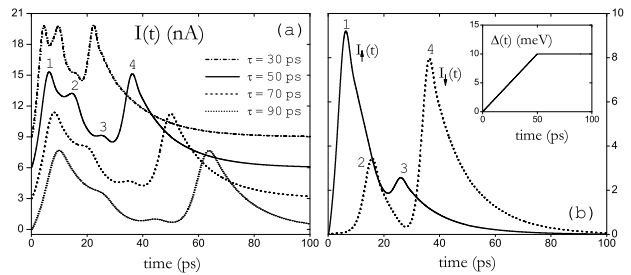}
\caption{\label{fig2} (a) Transient currents driven by a ramp--up
voltage. The lines represent different values of turn--on duration
$\tau$, and are separated at $t = 0$ by $3\,$nA for clarity. (b)
Transient current of each spin direction for $\tau = 50\,$ps. The
ramp--up voltage is depicted in the inset. Other parameters are (in
unit of meV): $\epsilon\dup = 0.5$, $\epsilon\ddn = 2.5$, $U = 4$,
$\Gamma = 0.05$, $W = 15$, $T = 0.2$, and $\Delta = 10$. }
\end{center}
\end{figure}

We now turn to cases of a finite switch--on duration
$0<\tau<\infty$, in which the lead level shift $\Delta(t)$ is
exemplified in the inset of~\Fig{fig2}(b). The response current
emerges at $t = 0$ immediately after the QD is excited out of
equilibrium, and then fluctuates as the lead chemical potential
${\mu}$ sweeps over the various QD Fock states. After $t = \tau$,
the current vanishes gradually as the reduced system is drawn
towards the steady state with ${\mu}(t>\tau) = \Delta$.
In~\Fig{fig2}(a) we plot the transient currents for various values
of $\tau$. Peaks associated with different resonances are observed
directly in the time domain, at its linear mapping to $\Delta(t)$
during the period of $t < \tau$. Especially, the calculated $I(t)$
for $\tau = 50\,$ps resolves distinctly four peaks labeled by
numbers from $1$ to $4$, which are responsible for all the four
resonances consecutively activated by the ramp--up voltage with an
increasing excitation energy of $\epsilon\dup$, $\epsilon\ddn$,
$\epsilon\dup + U$, and $\epsilon\ddn + U$, respectively.
Figure~\ref{fig2}(b) depicts $I_s(t)$ for both spins with $\tau =
50\,$ps. Within the period of $t < \tau$, the major peaks in
$I\dup(t)$ (peak--1) and $I\ddn(t)$ (peak--4) start to form as the
time--dependent $\mu(t)$ matches $\epsilon\dup$ and $\epsilon\ddn +
U$, respectively; and they remain discernible even for a rather
large $\tau$; see the line for $\tau = 90\,$ps in~\Fig{fig2}(a).
However, the amplitudes of the two minor peaks, peak--3 in
$I\dup(t)$ and peak--2 in $I\ddn(t)$, diminish rapidly as the
ramp--up duration $\tau$ is lengthened. In the adiabatic limit
($\tau \rightarrow \infty$), the minor peaks become completely
invisible, since the spin--up and spin--down QD--levels are charged
sequentially due to the Coulomb blockade; see~\Fig{fig1}(b).

\begin{figure}
\begin{center}
\includegraphics[width=0.95\columnwidth]{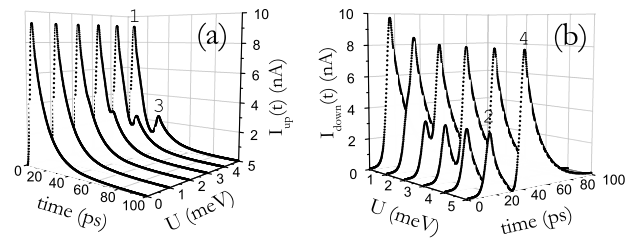}
\caption{ \label{fig3} (a) Spin--up and (b) spin--down transient
currents in response to a ramp--up voltage for different $U$. Same
parameters are adopted as in~\Fig{fig2}(b). }
\end{center}
\end{figure}

Influence of Coulomb interaction on the transient current is
illustrated by~\Fig{fig3}. Cases of different $U$ ranging from a
noninteracting scenario to a strongly blockaded QD are studied. For
$U=0$, there is only one peak in $I_s(t)$ for either spin, since the
excitation energies $\epsilon_s$ and $\epsilon_s + U$ are
indistinguishable. The peak is split into two with an increasing
displacement as $U$ is augmented. For either spin direction, the
first peak always sticks to its original position as in $U = 0$,
while the second one emerges later in time with a larger $U$. For
$U$ as large as $5\,$meV, the two peaks in $I\ddn(t)$ are almost
completely separated. This confirms our attribution of the peaks to
the various resonances: the first peaks in $I\dup(t)$ (peak--1) and
$I\ddn(t)$ (peak--2) correspond to the resonances ${\mu} =
\epsilon\dup$ and ${\mu} = \epsilon\ddn$, respectively; and the
second ones in $I\dup(t)$ (peak--3) and $I\ddn(t)$ (peak--4) are
associated with ${\mu} = \epsilon\dup + U$ and ${\mu} = \epsilon\ddn
+ U$, respectively.

\begin{figure}
\begin{center}
\includegraphics[width=0.95\columnwidth]{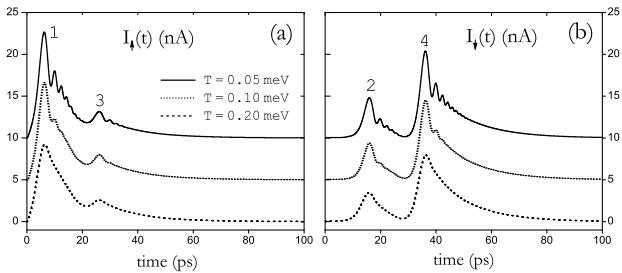}
\caption{ \label{fig4} Transient current for (a) spin--up and (b)
spin--down electrons in response to a ramp--up voltage with $\Delta
= 10\,$meV, $U = 4\,$meV, and $\tau = 50\,$ps. The lines represent
different temperatures, and separated at $t = 0$ by $5\,$nA for
clarity. Other parameters are same as in~\Fig{fig2}(b). }
\end{center}
\end{figure}

Thermal effect is also explored. $I_s(t)$ under different lead
temperatures are plotted in~\Fig{fig4}. A series of satellite
oscillations appears following each resonance peak in $I_s(t)$ at a
low $T$, which is resulted from a time--varying phase factor,
$\exp[\,i\! \int^t_\tau \! d\bar{t}\, \Delta(\bar{t})]$, of the
nonequilibrium lead correlation function. The oscillation frequency
is thus determined by the magnitude of $\Delta(t)$. For a ramp--up
voltage, $\Delta(t)$ increases monotonically during $0< t < \tau$.
It is thus reasonable to find the frequency of satellite
oscillations grows with time, and the decaying amplitude of the
oscillations is due to the dissipative interactions between the QD
and the semi--infinite lead.

To summarize, with a finite switch--on duration $\tau$, the QD
energetic configuration is resolved distinctly (sometimes
completely) through the resonance peaks of response current in
\emph{real time}, in obvious contrast to the adiabatic charging
limit.

\subsection{The instantaneous switch--on limit} \label{instant}

As $\tau$ approaches zero, a ramp--up voltage becomes asymptotically
a step pulse, \emph{i.e.}, $\Delta(t) = \Delta \Theta(t)$, with
$\Theta(t)$ being the Heaviside step function.
Upon the instantaneous switch--on of external voltage, both the
spin--up and down QD--levels are activated simultaneously, provided
the voltage amplitude $\Delta$ is sufficiently large. It is
intriguing to see in this limit how the response current would be
related to the QD energetic configuration.

\begin{figure}
\begin{center}
\includegraphics[width=0.95\columnwidth]{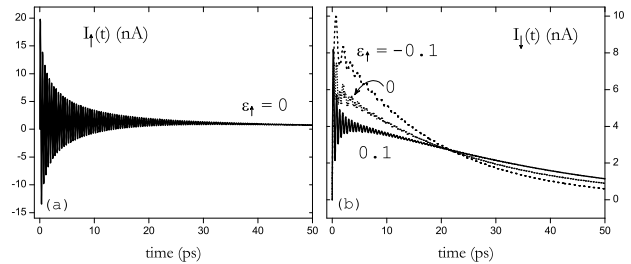}
\caption{\label{fig5} (a) $I\dup(t)$ with $\epsilon\dup=0$ and (b)
$I\ddn(t)$ with three values of $\epsilon\dup$ in unit of meV. Other
parameters are (in unit of meV): $\epsilon\ddn = 2.5$, $U = 4.5$, $W
= 5$, $\Gamma = 0.05$, $T = 0.01$, and $\Delta = 10$. }
\end{center}
\end{figure}

In \Fig{fig5} (a) and (b) we plot the calculated $I\dup(t)$ and
$I\ddn(t)$ in response to a step voltage of $\Delta = 10\,$meV,
respectively. Three QDs of different $\epsilon\dup$ are studied. The
energetics scenario for the three QDs are $\mu_{\rm eq} =
\epsilon\dup < \epsilon\ddn$, $\mu_{\rm eq} < \epsilon\dup <
\epsilon\ddn$, and $\epsilon\dup < \mu_{\rm eq}  < \epsilon\ddn$,
respectively. Note that for all cases are in a large applied voltage
regime since $\Delta > \epsilon_s$ and $\Delta > \epsilon_s + U$,
which ensure all the four QD Fock states accessible energetically.
For both spin directions, the transient current shows an initial
overshooting, and then oscillates on top of a decay as the reduced
system evolves into a steady state with the lead chemical potential
$\mu = \Delta$. These rapid oscillations are not artifacts. Their
presence reflects the electronic dynamics of QD driven by an
external voltage of large amplitude. Upon a small perturbation on
$\epsilon\dup$ of $0.1\,$meV, $I\dup(t)$ scarcely changes as those
with $\epsilon\dup=\pm\, 0.1\,$meV almost overlap with the
$\epsilon\dup=0$ case shown in \Fig{fig5}(a); while $I\ddn(t)$
displays drastic deviations among the three cases, in terms of the
overshooting amplitude, the oscillation frequency, as well as the
decaying rate.

\begin{figure}
\begin{center}
\includegraphics[width=0.95\columnwidth]{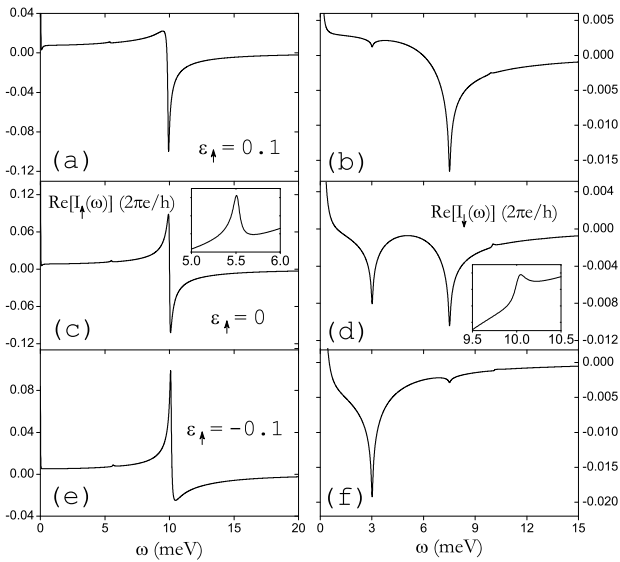}
\caption{\label{fig6} $\mbox{Re\,}[I\dup(\omega)]$ (left panels) and
$\mbox{Re\,}[I\ddn(\omega)]$ (right panels) for three specified
values of $\epsilon\dup$ (unit of meV) in the top, middle, and
bottom panels, respectively, in relation to $\mu_{\rm eq}$. Other
parameters are those adopted in \Fig{fig5}, and same for all panels.
Insets in (a) and (b) magnify the peaks at $5.5\,$meV and $10\,$meV,
respectively. }
\end{center}
\end{figure}

To understand the physical origin of the above observations,
frequency--dependent current spectrum is calculated, \emph{i.e.},
$I_s(\omega) \equiv \mathcal{F}[I_s(t)]$ with $\mathcal{F}$ denoting
the conventional Fourier transform. The results corresponding to the
three QDs are depicted in \Fig{fig6}. For all three cases, two
characteristic frequencies are revealed in
$\mbox{Re}[I\dup(\omega))]$, \emph{i.e.}, a major kink at $\Delta -
\epsilon\dup$ and a minor wiggle around $\Delta-\epsilon\dup-U$.
The line shape of $\mbox{Re\,}[I\dup(\omega)]$ around the frequency
$\omega\dup \equiv \Delta - \epsilon\dup$ much resembles that of a
single--level QD~\cite{Zheng08Paper3}. For $\epsilon\dup < \mu_{\rm eq}$,
$\mbox{Re\,}[I\dup(\omega)]$ exhibits a peak at $\omega\dup$;
while for $\epsilon\dup > \mu_{\rm eq}$, a dip shows up. In the case
of $\epsilon\dup = \mu_{\rm eq}$, $\mbox{Re\,}[I\dup(\omega)]$
around $\omega\dup$ is described by the function form of $\ln(x^2 +
1)/x$, with $x = 2(\omega - \omega\dup)/\Gamma$.
However, the profiles of $\mbox{Re}[I\ddn(\omega)]$ are
significantly different among the three cases. To be specific, for
$\epsilon\dup = \mu_{\rm eq}$, two prominent dips show up at
frequencies $\Delta-\epsilon\ddn$ and $\Delta-\epsilon\ddn-U$,
respectively; for $\epsilon\dup > \mu_{\rm eq}$, the dip at
$\Delta-\epsilon\ddn-U$ is considerably suppressed, which is
associated with the transition $\spup \leftrightarrow \spupdn$; and
for $\epsilon\dup < \mu_{\rm eq}$, it is the dip at
$\Delta-\epsilon$ that is almost completely quenched, which
indicates an inactive transition $\spzero \leftrightarrow \spdown$.
Hence it is demonstrated the response current spectrum varies
sensitively to the QD energetic configuration.

The presence (or absence) of a resonance signature in
$\mbox{Re}[I_s(\omega)]$, as well as the issue of peak versus dip at
$\omega\dup$, are closely related to the equilibrium occupancy of
the spin--up state prior to the switch--on of the voltage.
Specifically, in the case of $\epsilon\dup = 0.1\,{\rm meV}\,(>
\mu_{\rm eq})$, the initial population of spin--up electrons on the
QD is as small as $0.08$, and the subsequent incoming spin--down
electrons driven by the step voltage would feel scarcely any Coulomb
repulsion due to the almost vacant spin--up level. This thus leads
to the prominent dip at $\Delta-\epsilon\ddn$ and the barely
recognizable one at $\Delta-\epsilon\ddn-U$ in
$\mbox{Re}[I\ddn(\omega)]$; see \Fig{fig6}(b). Whereas in the case
of $\epsilon\dup = -0.1\,{\rm meV}\,(< \mu_{\rm eq})$, the spin--up
level is nearly fully occupied at $t=0$, \emph{i.e.}, $N\dup(0) =
0.92$. Therefore, the majority of spin--down electrons injected
afterwards need to acquire an excess energy of $U$ to overcome the
Coulomb blockade, and this is why the dip at $\Delta-\epsilon\ddn$
becomes trivial; see \Fig{fig6}(f). For $\epsilon\dup = \mu_{\rm
eq}$, the spin--up level is half occupied ($N\dup = 0.5$) prior to
the switch--on of voltage. In such a circumstance, both the resonant
transitions $\spzero \leftrightarrow \spdown$ and $\spup
\leftrightarrow \spupdn$ contribute equally to the response current,
and hence give rise to the two dips of similar depth at frequencies
$\Delta - \epsilon\ddn$ and $\Delta - \epsilon\ddn - U$; see
\Fig{fig6}(d). With $\epsilon\dup > 0.1\,$meV, the resonance
signature in $\mbox{Re}[I\dup(\omega)]$ at $\Delta-\epsilon\dup$
becomes a pure dip, and the minor dip in $\mbox{Re}[I\dup(\omega)]$
at $\Delta-\epsilon\ddn-U$ vanishes completely; while with
$\epsilon\dup <-0.1\,$meV, $\mbox{Re}[{I\dup(\omega)}]$ shows a pure
peak at $\Delta-\epsilon\dup$, and the minor dip in
$\mbox{Re}[I\dup(\omega)]$ at $\Delta-\epsilon\ddn$ becomes
invisible.

This thus confirms that in the instantaneous switch--on limit, the
$\omega$--dependent response current spectrum resolves sensitively
and faithfully the QD energetic configuration through the
characteristic resonance signatures.
To our surprise, $\mbox{Re\,}[I\ddn(\omega)]$ resolves a third peak
at $\omega\dup$; see the tiny bump in~\Fig{fig6}(b) and its inset.
It possibly arises due to the higher--order response, and the
mechanism of its presence requires further understanding on the
transient dynamic processes. At a lower temperature, all the
resonance signatures in the current spectrum are accentuated (not
shown), \emph{i.e.}, the peaks (dips) become higher (deeper) and
sharper. In the real time, this amounts to an enhanced oscillation
amplitude of transient current.

\section{Concluding remarks} \label{summary}

To conclude, we have investigated the transient electronic dynamics
of a single--lead interacting QD driven by a family of ramp--up
voltages. We have found that (a) in the adiabatic charging limit,
the quasi--steady dynamics provide only incomplete information on
the QD energetic configuration; (b) With an appropriately chosen
switch--on duration, all transitions among QD Fock states can be
resolved distinctly through the resonance peaks of transient current
in real time, and hence provide comprehensive (sometimes complete)
information on the QD energetic configuration; (c) In the
instantaneous switch--on limit, the frequency--dependent responses
current spectrum reveals sensitively and faithfully the QD energetic
configuration through the characteristic resonance signatures. This
work thus highlights the significance and versatility of transient
dynamics calculations.

In the sequential tunneling regime where $\Gamma \ll T$
(\emph{e.g.}, Figs.~\ref{fig1}$-$\ref{fig3}), the present HEOM with
truncation tier of $N_{\rm trun} = 1$,
 which amounts to
the conventional quantum master equation methods with
memory~\cite{Yan05187}, will be sufficient. When $\Gamma \sim T$,
calculations with $N_{\rm trun}=1$ would produce qualitatively
correct static and dynamic Coulomb blockade for a single--lead QD,
because the basic physics does not involve the cotunneling
mechanism. However, to obtain quantitatively accurate results,
$N_{\rm trun} = 2$ is needed. Extensive numerical tests indicate
that $N_{\rm trun}=2$ can support up to $\Gamma \sim 5T$
(\emph{e.g.}, Figs.~\ref{fig5} and \ref{fig6}). It has been shown
\cite{Jin08234703} that the present HEOM theory with $N_{\rm
trun}=2$, which properly treats the cotunneling dynamics, is
equivalent to a real--time diagrammatic formalism~\cite{Kon961715};
thus the Kondo problems can be addressed.
However, a higher truncation tier is numerically needed for
convergency when $T \ll \Gamma$.
The quantitatively accurate results presented in this work
(for $\Gamma < 5T$) are useful for further
development of efficient but maybe approximate computational
schemes.

Investigating and/or manipulating transient electronic dynamics
provide(s) sometimes a unique way to achieve physical properties and
novel functions of mesoscopic many--body systems.
However, exact theoretical results
for transient quantum transport are rare.
Weiss~\emph{et~al} \cite{Wei08195316} have recently proposed
an iterative real--time path integral approach,
which is numerically exact but expensive especially
when the applied bias voltage is beyond some simple forms such as
a step function.
The present HEOM theory  is formally exact and
provides an alternative approach. It is applicable
to arbitrary time--dependent external fields applied
to leads and/or to the system, without extra
computational cost. However, computational
effort increases dramatically if the converged results
require a high truncation tier of the
hierarchy~\cite{Jin08234703}.

\begin{acknowledgments}
  Support from the RGC (604007 and 604508) of Hong Kong is acknowledged.
\end{acknowledgments}



%

\end{document}